\documentclass{article}
\usepackage{amsmath, amssymb, amsfonts}
\title{IS THE RINDLER HORIZON ENERGY NONVANISHING ? \footnote{Essay selected for ''Honorable Mention'' in the 2006 Awards for Essays on Gravitation (Gravitation Research Foundation)}} 
\author{Hristu Culetu\\ Ovidius University, Dept.of Physics, B-dul Mamaia 124, \\ 8700 Constanta, Romania \\ email : hculetu@yahoo.com}
\begin{document}
\numberwithin{equation}{section}
\pagenumbering{arabic}
\maketitle
\begin{abstract}
A nonvanishing value for the Rindler horizon energy is proposed, by an analogy with the ''near horizon'' Schwarzschild metric. We show that the Rindler horizon energy is given by the same formula $E = \alpha/2$ obtained by Padmanabhan for the Schwarzschild spacetime, where $\alpha$ is the gravitational radius.\\
PACS : 04.70.Dy, 04.50.+h, 04.60.-m. 
\end{abstract}
\newcommand{\fv}{\boldsymbol{f}}
\newcommand{\tv}{\boldsymbol{t}}
\newcommand{\gv}{\boldsymbol{g}}
\newcommand{\OV}{\boldsymbol{O}}
\newcommand{\wv}{\boldsymbol{w}}
\newcommand{\WV}{\boldsymbol{W}}
\newcommand{\NV}{\boldsymbol{N}}
\newcommand{\hv}{\boldsymbol{h}}
\newcommand{\yv}{\boldsymbol{y}}
\newcommand{\RE}{\textrm{Re}}
\newcommand{\IM}{\textrm{Im}}
\newcommand{\rot}{\textrm{rot}}
\newcommand{\dv}{\boldsymbol{d}}
\newcommand{\grad}{\textrm{grad}}
\newcommand{\Tr}{\textrm{Tr}}
\newcommand{\ua}{\uparrow}
\newcommand{\da}{\downarrow}
\newcommand{\ct}{\textrm{const}}
\newcommand{\xv}{\boldsymbol{x}}
\newcommand{\mv}{\boldsymbol{m}}
\newcommand{\rv}{\boldsymbol{r}}
\newcommand{\kv}{\boldsymbol{k}}
\newcommand{\VE}{\boldsymbol{V}}
\newcommand{\sv}{\boldsymbol{s}}
\newcommand{\RV}{\boldsymbol{R}}
\newcommand{\pv}{\boldsymbol{p}}
\newcommand{\PV}{\boldsymbol{P}}
\newcommand{\EV}{\boldsymbol{E}}
\newcommand{\DV}{\boldsymbol{D}}
\newcommand{\BV}{\boldsymbol{B}}
\newcommand{\HV}{\boldsymbol{H}}
\newcommand{\MV}{\boldsymbol{M}}
\newcommand{\be}{\begin{equation}}
\newcommand{\ee}{\end{equation}}
\newcommand{\ba}{\begin{eqnarray}}
\newcommand{\ea}{\end{eqnarray}}
\newcommand{\bq}{\begin{eqnarray*}}
\newcommand{\eq}{\end{eqnarray*}}
\newcommand{\pa}{\partial}
\newcommand{\f}{\frac}
\newcommand{\FV}{\boldsymbol{F}}
\newcommand{\ve}{\boldsymbol{v}}
\newcommand{\AV}{\boldsymbol{A}}
\newcommand{\jv}{\boldsymbol{j}}
\newcommand{\LV}{\boldsymbol{L}}
\newcommand{\SV}{\boldsymbol{S}}
\newcommand{\av}{\boldsymbol{a}}
\newcommand{\qv}{\boldsymbol{q}}
\newcommand{\QV}{\boldsymbol{Q}}
\newcommand{\ev}{\boldsymbol{e}}
\newcommand{\uv}{\boldsymbol{u}}
\newcommand{\KV}{\boldsymbol{K}}
\newcommand{\ro}{\boldsymbol{\rho}}
\newcommand{\si}{\boldsymbol{\sigma}}
\newcommand{\thv}{\boldsymbol{\theta}}
\newcommand{\bv}{\boldsymbol{b}}
\newcommand{\JV}{\boldsymbol{J}}
\newcommand{\nv}{\boldsymbol{n}}
\newcommand{\lv}{\boldsymbol{l}}
\newcommand{\om}{\boldsymbol{\omega}}
\newcommand{\Om}{\boldsymbol{\Omega}}
\newcommand{\Piv}{\boldsymbol{\Pi}}
\newcommand{\UV}{\boldsymbol{U}}
\newcommand{\iv}{\boldsymbol{i}}
\newcommand{\nuv}{\boldsymbol{\nu}}
\newcommand{\muv}{\boldsymbol{\mu}}
\newcommand{\lm}{\boldsymbol{\lambda}}
\newcommand{\Lm}{\boldsymbol{\Lambda}}
\newcommand{\opsi}{\overline{\psi}}
\renewcommand{\tan}{\textrm{tg}}
\renewcommand{\cot}{\textrm{ctg}}
\renewcommand{\sinh}{\textrm{sh}}
\renewcommand{\cosh}{\textrm{ch}}
\renewcommand{\tanh}{\textrm{th}}
\renewcommand{\coth}{\textrm{cth}}

~~One of the basic features of classical gravity is the generation of surfaces acting as one - way membranes. The typical example is given by the Schwarzschild black hole which has a closed observer - independent surface (event horizon). 
~The de Sitter spacetime has also a compact surface which is however dependent upon the observer. A one way membrane may be induced even in flat spacetime, with the appearance of an observer - dependent horizon (the non - compact Rindler horizon).

The problem is whether one could associate thermodynamic parameters (temperature, entropy and energy) to that horizon, as for the black hole. One of the authors who developed a model on the subject was T. Padmanabhan \cite {TP1}. He found that
\begin{equation}
S = \frac{1}{4} 4 \pi \alpha^{2}= \frac{A_{hor}}{4} ; ~~~~E = \frac{\alpha}{2}
\label{1}
\end{equation}
where $S$ and $E$ are the entropy and energy of the Schwarzschild black hole and $A_{hor}$ -~the horizon area, of radius $\alpha$. For a spacetime with planar symmetry (for instance, the Rindler one), he established that $S = (1/4) A_{\bot}$ while the energy is vanishing ($A_{\bot}$ is a finite part of the infinite transverse area).

We try in this paper to prove that the expression $E = \alpha/2$ is valid for the Rindler spacetime, too.

F.Alexander and U.Gerlach \cite{AG} showed that the electric field of a uniformly accelerated charge $e$ induces on the event horizon a surface charge density such that the total charge on that surface is $-e$. Moreover, an attractive force between the point charge and the horizon appears, its expression proving to be equal to the radiation reaction force from the r.h.s. of the Lorentz - Dirac equation.

 By analogy with the electromagnetic case, we suggest that the Rindler horizon contains an (observer dependent) surface density $\sigma$ given by \cite {MP}
\begin{equation}
g = 4 \pi G \sigma.
\label{2}
\end{equation}
where g is the proper acceleration of the hyperbolic observer.\\
From now on we take $G = \hbar = c = k_{B} = 1$.

~To show that $E = \alpha/2$ is true also for the Rindler horizon, we look for an analogy with Schwarzschild's geometry.

It is well known that near the horizon $r = 2M$ of the black hole, the Schwarzschild line element may be written in the form
\begin{equation}
ds^{2} =  -\frac{r-2M}{2M} dt^{2}+\frac{2M}{r-2M} dr^{2}+ dL_{\bot}^{2}
\label{3}
\end{equation}
where $dL_{\bot}^{2}$ stands for the transverse coordinates.
~By means of the transformation $\rho = 2 \sqrt{2M~(r-2M)}$, eq. (3) becomes
\begin{equation}
ds^{2} = -(\frac{1}{4M})^{2} \rho^{2} dt^{2}+d \rho^{2}+dL_{\bot}^{2}
\label{4}
\end{equation}
which is the Rindler spacetime.\\
~~It is clear that the surface gravity $\kappa = 1/4M$ of the black hole plays a similar role with the proper acceleration $g$ of the uniformly accelerated observer. Keeping in mind that the energy of the black hole is given by $E = M = 1/4 \kappa$, we have therefore for the Rindler horizon
\begin{equation}
E(g) = \frac{1}{4g} 
\label{5}
\end{equation}
which gives again $E = \alpha/2$. We mention that Kerner and Mann \cite {KM} reach the same result using the Hamilton - Jacobi ansatz method.

We have, of course, tacitly assumed that all the energy of the black hole is located at the horizon $r = \alpha$. A similar situation is faced when the Vilenkin - Ipser - Sikivie (VIS) \cite{IS} \cite{AV} domain wall is studied. As Chamblin and Eardley \cite{CE} have noticed, ``we think of a VIS spacetime as an inflating universe where all of the vacuum energy has been concentrated on the sheet of the domain wall``. \\
~With all fundamental constants, eq. (5) appears as $E = (c^{4}/4G)~(c^{2}/g)$.

 The surface energy density on the horizon looks now as 
\begin{equation}
\sigma = \frac{1}{4g} \frac{1}{4 \pi (4M^{2})} = \frac{g}{4 \pi}
\label{6}
\end{equation}
whence $g = 4 \pi \sigma$, as expected (for instance, from Gauss' theorem \cite {MP}).

Having established that the energy of the ``horizon membrane`` is given by (5), we pass now to the thermodynamical parameters. The temperature is, of course, the Unruh temperature $T_{U} = g/2 \pi$ (it is well known that the Davies - Unruh radiation comes from the horizon of the accelerated observer).

 For the entropy of the Rindler horizon one obtains $S(g) = 4 \pi M^{2} = \pi/4g^{2}$.
~It is easy to check that the thermodynamic relation $dE = T dS$ is obeyed.\\
~Even though we started with an analogy between the Schwarzschild and Rindler spacetimes near the black hole horizon, we believe the Rindler horizon contains a surface energy given by eq. (6) and generated by the agent who accelerates the test particle. We also stress that eq. (5) preserves its form even for the planar horizon of an uniformly accelerated observer, in Cartesian coordinates. This might be justified basing on the fact that, for an observer located near the black hole horizon, it appears as a flat surface (the linear dimension of the observer is considered to be much less than the Schwarzschild radius).

It is worth to notice a relation between the classical expression (5) for the Rindler energy and the quantum one \cite{HC} $E_{quan} = g/2$. 
Since $E_{class} \propto 1/g$ and $E_{quan} \propto g$, we have 
\begin{equation}
 E_{quan}~ E_{class} \approx \epsilon_{P}^{2}.
\label{7}
\end{equation}
~In other words, the Planck energy $\epsilon_{P}$ is of the order of the geometrical mean of the two, playing the role of a ``boundary`` between the two domains of energies.

To summarize, we showed in this letter that Padmanabhan's expression $E = |\alpha/2|$ for the Schwarzschild and de Sitter horizon energies is valid also for the Rindler horizon. In addition, the corresponding entropy $S(g)$ is finite and the thermodynamic relation $dE = T dS$ is fulfilled.\\

\end{document}